\documentclass[5p,preprint,longtitle,times,twocolumn]{elsarticle} 

\usepackage{lineno}
\usepackage{amsmath,amsfonts,amsthm,amsbsy}
\usepackage{amssymb}
\usepackage{url}

\usepackage{breakurl}
\usepackage{hyperref}

\modulolinenumbers[1]

\journal{Physics Letters B}

\begin{document}

\begin{frontmatter}

\title{Total and differential cross sections of $\eta$-production in proton-deuteron fusion for excess energies between $Q_\eta=13\;\text{MeV}$ and $Q_\eta=81\;\text{MeV}$} 



\author[IKPUU]{The WASA-at-COSY Collaboration\\[2ex] P.~Adlarson}
\author[ASWarsN]{W.~Augustyniak}
\author[IPJ]{W.~Bardan}
\author[Edinb]{M.~Bashkanov}
\author[MS]{F.S.~Bergmann}
\author[ASWarsH]{M.~Ber{\l}owski}
\author[Budker,Novosib]{A.~Bondar}
\author[PGI,DUS]{M.~B\"uscher}
\author[IKPUU]{H.~Cal\'{e}n}
\author[IFJ]{I.~Ciepa{\l}}
\author[PITue,Kepler]{H.~Clement}
\author[IPJ]{E.~Czerwi{\'n}ski}
\author[MS]{K.~Demmich}
\author[IKPJ]{R.~Engels}
\author[ZELJ]{A.~Erven}
\author[ZELJ]{W.~Erven}
\author[Erl]{W.~Eyrich}
\author[IKPJ,ITEP]{P.~Fedorets}
\author[Giess]{K.~F\"ohl}
\author[IKPUU]{K.~Fransson}
\author[IKPJ]{F.~Goldenbaum}
\author[IKPJ,IITI]{A.~Goswami}
\author[IKPJ,HepGat]{K.~Grigoryev}
\author[IKPUU]{C.--O.~Gullstr\"om}
\author[IKPJ,IASJ]{C.~Hanhart}
\author[IKPUU]{L.~Heijkenskj\"old\fnref{fnmz}}
\author[IKPJ]{V.~Hejny}
\author[MS]{N.~H\"usken\corref{coau}}\ead{n\_hues02@uni-muenster.de}
\author[IPJ]{L.~Jarczyk}
\author[IKPUU]{T.~Johansson}
\author[IPJ]{B.~Kamys}
\author[ZELJ]{G.~Kemmerling\fnref{fnjcns}}
\author[IPJ]{G.~Khatri\fnref{fnharv}}
\author[MS]{A.~Khoukaz}
\author[IPJ]{A.~Khreptak}
\author[HeJINR]{D.A.~Kirillov}
\author[IPJ]{S.~Kistryn}
\author[ZELJ]{H.~Kleines\fnref{fnjcns}}
\author[Katow]{B.~K{\l}os}
\author[ASWarsH]{W.~Krzemie{\'n}}
\author[IFJ]{P.~Kulessa}
\author[IKPUU,ASWarsH]{A.~Kup{\'s}{\'c}}
\author[Budker,Novosib]{A.~Kuzmin}
\author[NITJ]{K.~Lalwani}
\author[IKPJ]{D.~Lersch}
\author[IKPJ]{B.~Lorentz}
\author[IPJ]{A.~Magiera}
\author[IKPJ,JARA]{R.~Maier}
\author[IKPUU]{P.~Marciniewski}
\author[ASWarsN]{B.~Maria{\'n}ski}
\author[ASWarsN]{H.--P.~Morsch}
\author[IPJ]{P.~Moskal}
\author[IKPJ]{H.~Ohm}
\author[IFJ]{W.~Parol}
\author[PITue,Kepler]{E.~Perez del Rio\fnref{fnlnf}}
\author[HeJINR]{N.M.~Piskunov}
\author[IKPJ]{D.~Prasuhn}
\author[IKPUU,ASWarsH]{D.~Pszczel}
\author[IFJ]{K.~Pysz}
\author[IKPUU,IPJ]{A.~Pyszniak}
\author[IKPJ,JARA,Bochum]{J.~Ritman}
\author[IITI]{A.~Roy}
\author[IPJ]{Z.~Rudy}
\author[IPJ]{O.~Rundel}
\author[IITB]{S.~Sawant}
\author[IKPJ]{S.~Schadmand}
\author[IPJ]{I.~Sch\"atti--Ozerianska}
\author[IKPJ]{T.~Sefzick}
\author[IKPJ]{V.~Serdyuk}
\author[Budker,Novosib]{B.~Shwartz}
\author[MS]{K.~Sitterberg}
\author[PITue,Kepler,Tomsk]{T.~Skorodko}
\author[IPJ]{M.~Skurzok}
\author[IPJ]{J.~Smyrski}
\author[ITEP]{V.~Sopov}
\author[IKPJ]{R.~Stassen}
\author[ASWarsH]{J.~Stepaniak}
\author[Katow]{E.~Stephan}
\author[IKPJ]{G.~Sterzenbach}
\author[IKPJ]{H.~Stockhorst}
\author[IKPJ,JARA]{H.~Str\"oher}
\author[IFJ]{A.~Szczurek}
\author[ASWarsN]{A.~Trzci{\'n}ski}
\author[IKPUU]{M.~Wolke}
\author[IPJ]{A.~Wro{\'n}ska}
\author[ZELJ]{P.~W\"ustner}
\author[KEK]{A.~Yamamoto}
\author[ASLodz]{J.~Zabierowski}
\author[IPJ]{M.J.~Zieli{\'n}ski}
\author[IKPUU]{J.~Z{\l}oma{\'n}czuk}
\author[ASWarsN]{P.~{\.Z}upra{\'n}ski}
\author[IKPJ]{M.~{\.Z}urek}

\author[Colin]{\vskip1em and\vskip1em C.~Wilkin}

\address[IKPUU]{Division of Nuclear Physics, Department of Physics and 
 Astronomy, Uppsala University, Box 516, 75120 Uppsala, Sweden}
\address[ASWarsN]{Department of Nuclear Physics, National Centre for Nuclear 
 Research, ul.\ Hoza~69, 00-681, Warsaw, Poland}
\address[IPJ]{Institute of Physics, Jagiellonian University, prof.\ 
 Stanis{\l}awa {\L}ojasiewicza~11, 30-348 Krak\'{o}w, Poland}
\address[Edinb]{School of Physics and Astronomy, University of Edinburgh, 
 James Clerk Maxwell Building, Peter Guthrie Tait Road, Edinburgh EH9 3FD, 
 Great Britain}
\address[MS]{Institut f\"ur Kernphysik, Westf\"alische Wilhelms--Universit\"at 
 M\"unster, Wilhelm--Klemm--Str.~9, 48149 M\"unster, Germany}
\address[ASWarsH]{High Energy Physics Department, National Centre for Nuclear 
 Research, ul.\ Hoza~69, 00-681, Warsaw, Poland}
\address[Budker]{Budker Institute of Nuclear Physics of SB RAS, 11~akademika 
 Lavrentieva prospect, Novosibirsk, 630090, Russia}
\address[Novosib]{Novosibirsk State University, 2~Pirogova Str., Novosibirsk, 
 630090, Russia}
\address[PGI]{Peter Gr\"unberg Institut, PGI--6 Elektronische Eigenschaften, 
 Forschungszentrum J\"ulich, 52425 J\"ulich, Germany}
\address[DUS]{Institut f\"ur Laser-- und Plasmaphysik, Heinrich--Heine 
 Universit\"at D\"usseldorf, Universit\"atsstr.~1, 40225 D\"usseldorf, Germany}
\address[IFJ]{The Henryk Niewodnicza{\'n}ski Institute of Nuclear Physics, 
 Polish Academy of Sciences, Radzikowskiego~152, 31--342 Krak\'{o}w, Poland}
\address[PITue]{Physikalisches Institut, Eberhard--Karls--Universit\"at 
 T\"ubingen, Auf der Morgenstelle~14, 72076 T\"ubingen, Germany}
\address[Kepler]{Kepler Center f\"ur Astro-- und Teilchenphysik, 
 Physikalisches Institut der Universit\"at T\"ubingen, Auf der 
 Morgenstelle~14, 72076 T\"ubingen, Germany}
\address[IKPJ]{Institut f\"ur Kernphysik, Forschungszentrum J\"ulich, 52425 
 J\"ulich, Germany}
\address[ZELJ]{Zentralinstitut f\"ur Engineering, Elektronik und Analytik, 
 Forschungszentrum J\"ulich, 52425 J\"ulich, Germany}
\address[Erl]{Physikalisches Institut, Friedrich--Alexander--Universit\"at 
 Erlangen--N\"urnberg, Erwin--Rommel-Str.~1, 91058 Erlangen, Germany}
\address[ITEP]{Institute for Theoretical and Experimental Physics named 
 by A.I.\ Alikhanov of National Research Centre ``Kurchatov Institute'', 
 25~Bolshaya Cheremushkinskaya, Moscow, 117218, Russia}
\address[Giess]{II.\ Physikalisches Institut, Justus--Liebig--Universit\"at 
 Gie{\ss}en, Heinrich--Buff--Ring~16, 35392 Giessen, Germany}
\address[IITI]{Department of Physics, Indian Institute of Technology Indore, 
 Khandwa Road, Simrol, Indore - 453552, Madhya Pradesh, India}
\address[HepGat]{High Energy Physics Division, Petersburg Nuclear Physics 
 Institute named by B.P.\ Konstantinov of National Research Centre ``Kurchatov 
 Institute'', 1~mkr.\ Orlova roshcha, Leningradskaya Oblast, Gatchina, 188300, 
 Russia}
\address[IASJ]{Institute for Advanced Simulation, Forschungszentrum J\"ulich, 
 52425 J\"ulich, Germany}
\address[HeJINR]{Veksler and Baldin Laboratory of High Energiy Physics, 
 Joint Institute for Nuclear Physics, 6~Joliot--Curie, Dubna, 141980, Russia}
\address[Katow]{August Che{\l}kowski Institute of Physics, University of 
 Silesia, Uniwersytecka~4, 40--007, Katowice, Poland}
\address[NITJ]{Department of Physics, Malaviya National Institute of 
 Technology Jaipur, JLN Marg Jaipur - 302017, Rajasthan, India}
\address[JARA]{JARA--FAME, J\"ulich Aachen Research Alliance, Forschungszentrum
 J\"ulich, 52425 J\"ulich, and RWTH Aachen, 52056 Aachen, Germany}
\address[Bochum]{Institut f\"ur Experimentalphysik I, Ruhr--Universit\"at 
 Bochum, Universit\"atsstr.~150, 44780 Bochum, Germany}
\address[IITB]{Department of Physics, Indian Institute of Technology Bombay, 
 Powai, Mumbai - 400076, Maharashtra, India}
\address[Tomsk]{Department of Physics, Tomsk State University, 36~Lenina 
 Avenue, Tomsk, 634050, Russia}
\address[KEK]{High Energy Accelerator Research Organisation KEK, Tsukuba, 
 Ibaraki 305--0801, Japan}
\address[ASLodz]{Astrophysics Division, National Centre for Nuclear Research, 
 Box~447, 90--950 {\L}\'{o}d\'{z}, Poland}
\address[Colin]{Physics and Astronomy Department, UCL, Gower Street, London WC1E 6BT, United Kingdom}

\fntext[fnmz]{present address: Institut f\"ur Kernphysik, Johannes 
 Gutenberg--Universit\"at Mainz, Johann--Joachim--Becher Weg~45, 55128 Mainz, 
 Germany}
\fntext[fnjcns]{present address: J\"ulich Centre for Neutron Science JCNS, 
 Forschungszentrum J\"ulich, 52425 J\"ulich, Germany}
\fntext[fnharv]{present address: Department of Physics, Harvard University, 
 17~Oxford St., Cambridge, MA~02138, USA}
\fntext[fnlnf]{present address: INFN, Laboratori Nazionali di Frascati, Via 
 E.~Fermi, 40, 00044 Frascati (Roma), Italy}

\cortext[coau]{Corresponding author}

\begin{abstract}
New data on both total and differential cross sections of the production of $\eta$ mesons in proton-deuteron fusion to ${}^3\text{He}\,\eta$ in the excess energy region $13.6\;\text{MeV}\leq Q_\eta \leq 80.9\;\text{MeV}$ are presented. These data have been obtained with the WASA-at-COSY detector setup located at the Forschungszentrum J\"ulich, using a proton beam at 15 different beam momenta between $p_p = 1.60\;\text{GeV}/c$ and $p_p = 1.74\;\text{GeV}/c$. While significant structure of the total cross section is observed in the energy region $20\;\text{MeV}\lesssim Q_\eta \lesssim 60\;\text{MeV}$, a previously reported sharp variation around $Q_\eta\approx 50\;\text{MeV}$ cannot be confirmed. Angular distributions show the typical forward-peaking that was reported elsewhere. For the first time, it is possible to study the development of these angular distributions with rising excess energy over a large interval. 
\end{abstract}


\end{frontmatter}


\section{Introduction}

The production of $\eta$ mesons off nuclei has been a topic of active research over at least two decades. Inspired by the attractive interaction between $\eta$ mesons and nuclei, first studied by Bhalerao, Haider and Liu \cite{Bhalerao:1985cr, HAIDER1986257}, extensive experimental effort was put into the study of near-threshold production of $\eta$ mesons off various nuclei \cite{Chrien:1988gn, Fujioka:2012zz, Johnson:1993zy, Afanasiev:2011zz, Budzanowski:2008fr, Moskal:2010ee, Adlarson:2013xg, Adlarson:2016dme}. Although the original work suggested studies on heavier nuclei, the reaction $pd\rightarrow {}^3\text{He}\,\eta$ is one of the most discussed due to its markedly enhanced cross section very close to the production threshold. Here, it was observed that the production cross section $\sigma$ rises from zero at threshold to around $400\;\text{nb}$ within less than $1\;\text{MeV}$ of excess energy \cite{Berger:1988ba, Mayer:1995nu, Smyrski:2007nu, Mersmann:2007gw}. This curious behaviour of the production cross section has first been discussed in the context of a strong final state interaction and the presence of a possible (quasi-)bound $\eta{}^3\text{He}$ state close to the threshold in \cite{Wilkin:1993fe}, which was later followed up on, e.g., in \cite{WILKIN200792, Xie:2016zhs}. However, while the production cross section of the reaction $pd\rightarrow {}^3\text{He}\,\eta$ has been studied in great detail close to threshold, at higher excess energies the available database becomes sparse. Measurements by the CELSIUS/WASA \cite{Bilger:2002aw}, COSY-11 \cite{Adam:2007gz} and ANKE experiments \cite{Rausmann:2009dn} seem to suggest a cross section plateau away from threshold, whereas a measurement by the GEM collaboration \cite{Betigeri:1999qa} yielded a larger cross section value, albeit with a sizable uncertainty. Recently, in \cite{Adlarson:2014ysb}, a sharp variation of the total cross section between $Q_\eta = 48.8\;\text{MeV}$ and $Q_\eta = 59.8\;\text{MeV}$ has been reported. In order to further investigate the existence and cause of this cross section variation, a new measurement was performed at 15 different beam momenta between $p_p = 1.60\;\text{GeV}/c$ and $p_p = 1.74\;\text{GeV}/c$, using the experimental apparatus WASA at the COoler SYnchrotron COSY. Apart from determining the total cross section of the proton-deuteron fusion to the ${}^3\text{He}\,\eta$ final state, the focus of the new measurement is on the precise determination of differential cross sections and the study of their development with rising excess energy. Such a comparison between differential distributions at different excess energies has thus far been hindered by large systematic differences between the individual measurements performed in the various experiments mentioned above. For this reason, a coherent measurement over a large range of higher excess energies by a single experiment does for the first time present the possibility for an in-depth study of the dependence of the differential cross section on the excess energy. Here, high quality data are of great importance in order to facilitate theoretical work on the production mechanism of $\eta$ mesons in proton-deuteron fusion, as has recently been claimed in \cite{Kelkar:2013lwa}. Up to now, no model exists that manages to correctly reproduce the total and differential cross sections away from the production threshold. While the two-step model, first studied by \cite{Kilian:1990kh} in a classical framework and by \cite{FALDT1995769} quantum-mechanically, has some success in describing near-threshold data (see, e.g., \cite{Kelkar:2013lwa, FALDT1995769}), at larger excess energies the model no longer describes the available database \cite{Khemchandani:2007ta}. In \cite{Santra:2001ub}, it was reported that the GEM data can be adequately described by a resonance model, in which $\eta$ mesons are produced from the decay of a $N^*$ resonance. Such a model is, however, unlikely to have a large contribution close to threshold due to the large momentum transfer necessary to compensate for the $\eta$ meson mass. It remains to be resolved if, why, and at which energy the production mechanism of the reaction $pd\rightarrow {}^3\text{He}\,\eta$ changes. It is for these reasons that in \cite{Kelkar:2013lwa} new data at larger excess energies were assessed to be of high importance.

\section{Experiment}

The measurement was performed using the WASA detector setup (which is described in detail in \cite{wasa}) at the storage ring COSY of the Forschungszentrum J\"ulich. Utilizing the so-called supercycle mode of the storage ring, the momenta of the beam protons are changed at each injection of a new proton bunch. Eight beam settings can be stored at once and the measurement is composed of two such supercycles (SC), each containing the eight beam momenta (flat-tops) indicated in table \ref{tab:sc}. In total, data were taken at 15 different beam momenta between $p_p = 1.60\;\text{GeV}/c$ and $p_p = 1.74\;\text{GeV}/c$ with a momentum spread of around $\Delta p/p=10^{-3}$ \cite{Maier:1997zj} and a stepsize of $10\;\text{MeV}/c$. The measurement at a momentum of $p_p = 1.70\; \text{GeV}/c$ was repeated during both supercycles and in an additional single-energy measurement for systematic checks.
\begin{table}[!ht]
\centering
\caption{Nominal beam momenta $p_p$ for each supercycle and flat-top in $\text{GeV}/c$.}
\resizebox{0.95\textwidth}{!}{\begin{minipage}{\textwidth}
\label{tab:sc}
\begin{tabular}{l|llllllll}
\hline
  & FT0 & FT1 & FT2 & FT3 & FT4 & FT5 & FT6 & FT7 \\ 
\hline SC0 & $1.60$ & $1.62$ & $1.64$ & $1.66$ & $1.68$ & $1.70$ & $1.72$ & $1.74$ \\ 
 SC1 &  $1.61$ & $1.63$ & $1.65$ & $1.67$ & $1.69$ & $1.70$ & $1.71$ & $1.73$  \\ 
 SC2 & & & & & & $1.70$ & & \\
\hline
\end{tabular} 
\end{minipage}}
\end{table}
Inside the WASA Central Detector the beam protons are steered to collide with a deuterium pellet target. Due to the fixed-target geometry, heavy ejectiles like ${}^3\text{He}$ are produced near the forward direction and subsequently stopped inside the WASA Forward Detector. Here, using a proportional chamber and various layers of plastic scintillator, both the production angles $\vartheta$ and $\varphi$, and the energy deposit of forward-going particles are reconstructed. Doubly charged Helium ions can be efficiently separated from protons, deuterons and charged pions by their energy deposit. From the deposited energy, the kinetic energy of ${}^3\text{He}$ nuclei is also reconstructed, thus, in combination with the determined scattering angles, fully reconstructing their four-momenta.

\section{Data Analysis}

\begin{figure}[!ht]
	\centering
	\includegraphics[width=0.45\textwidth]{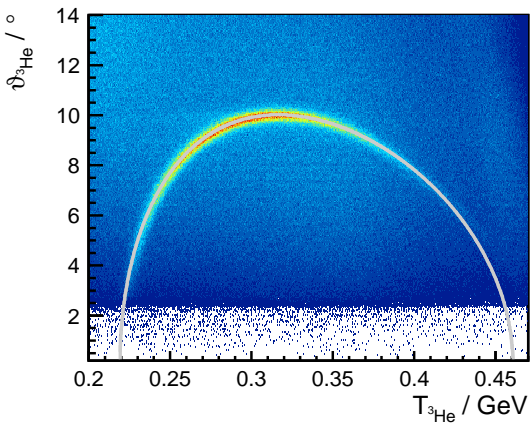}
	\caption{Distribution of polar angle $\vartheta_{{}^3\text{He}}$ versus kinetic energy $T_{{}^3\text{He}}$ of ${}^3\text{He}$ candidates stopped in the first layer of the WASA Forward Range Hodoscope from the measurement at $p_p = 1.70\; \text{GeV}/c$. The grey line shows the kinematical expectation for the reaction $pd\rightarrow {}^3\text{He}\,\eta$ at $p_p = 1.70\; \text{GeV}/c$, whereas the colour of the histograms reflects the number of reconstructed ${}^3\text{He}$ nuclei.}
	\label{fig:ekintheta}	
\end{figure}

\begin{figure}[!ht]
	\centering
		\includegraphics[width=0.5\textwidth]{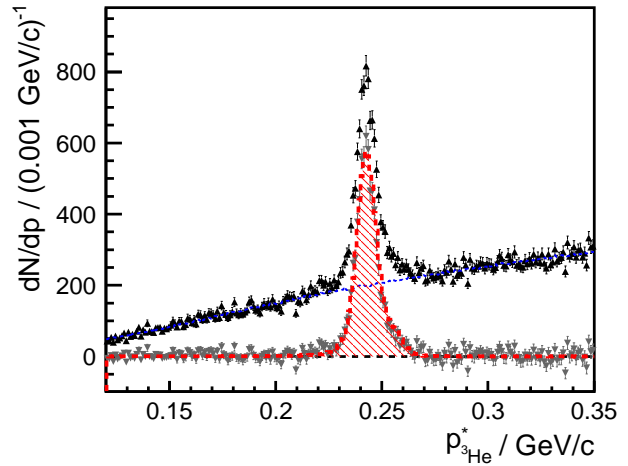}
	\caption{Example of a background fit to the final state momentum spectrum of ${}^3\text{He}$ nuclei for $0.5\leq \cos\vartheta_{\eta}^*<0.52$ at $p_p = 1.70\; \text{GeV}/c$. Black triangles with black error bars represent measured data, the blue dashed line represents the estimated background, gray downward triangles with gray error bars show the same data, subtracted by the background expectation. The red shaded histogram shows a MC simulation of the signal reaction $pd\rightarrow {}^3\text{He}\,\eta$.}
	\label{fig:examplefit-eta}	
\end{figure}

\begin{figure}[!ht]
	\centering
		\includegraphics[width=0.5\textwidth]{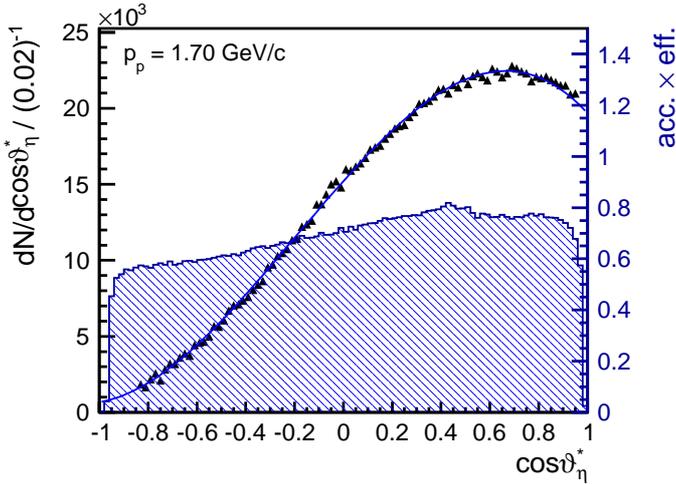}
	\caption{Angular distribution of the reaction $pd\rightarrow {}^3\text{He}\,\eta$ at $p_p = 1.70\; \text{GeV}/c$. Black triangles represent data, the blue line a polynomial fit of the kind given in Eq.(\ref{eq:pol}). The blue shaded histogram displays the corresponding product of acceptance and reconstruction efficiency in each bin in $\cos\vartheta_{\eta}^*$, with the axis displayed on the right side of the figure. Only statistical uncertainties are shown.}
	\label{fig:angdist-1700}	
\end{figure}

For a two-particle final state such as ${}^3\text{He}\,\eta$, the polar angle $\vartheta_{{}^3\text{He}}$ and the kinetic energy $T_{{}^3\text{He}}$ of the Helium nuclei are kinematically correlated. Using this relation, the precise measurement of the polar angle $\vartheta_{{}^3\text{He}}$ ($\Delta\vartheta_{{}^3\text{He}}\approx 0.2^\circ$) can be exploited to find a highly accurate calibration of the reconstructed energy. A comparison of the two-dimensional distribution of $\vartheta_{{}^3\text{He}}$ versus $T_{{}^3\text{He}}$ between the kinematical expectation for the signal reaction $pd\rightarrow {}^3\text{He}\,\eta$ and the data obtained at $p_p = 1.70\; \text{GeV}/c$ can be found in Fig. \ref{fig:ekintheta}.
The reaction of interest is identified from the spectra of the final state momentum of ${}^3\text{He}$ nuclei in the centre-of-mass frame $p_{{}^3\text{He}}^*$ in a missing-mass analysis. Thus, no assumption on the $\eta$ decay is made. Dividing the cosine of the centre-of-mass scattering angle $\cos\vartheta_{\eta}^*$ into 100 equally sized bins, the final state momentum spectra are fitted by a background function, excluding the peak region. Here, the background is a sum of Monte Carlo (MC) simulations of two- and three-pion production as well as a third order polynomial, accounting both for other possible background reactions and deviations from simple phase space distributions in the case of the three-pion production. The simulation of double-pion production was performed using a model incorporating the ABC effect and t-channel double-$\Delta(1238)$ excitation, developed for \cite{Adlarson:2014xmp}. An example of such a fit can be found in Fig. \ref{fig:examplefit-eta}.
In order to determine the signal yield in a given bin in $\cos\vartheta_{\eta}^*$, the background subtracted data are summed over the interval $p_\eta^*-3\sigma\leq p_{{}^3\text{He}}^*\leq p_\eta^*+3\sigma$, where $p_\eta^*$ and $\sigma$ are the position and width of the signal peak determined from a fit of an appropriate peak function to the background subtracted data. For most values of $\cos\vartheta_{\eta}^*$ a simple Gaussian is chosen. However, close to the maximum scattering angle the breakup of ${}^3\text{He}$ nuclei in the detector leads to asymmetric peaks (see Fig \ref{fig:examplefit-eta}) that are fitted by a double-Gaussian. In these cases, peak position and width of the dominant signal contribution are used. Before physically meaningful angular distributions are obtained, the signal yield needs to be corrected for the product of detector acceptance and reconstruction efficiency. This can be derived from MC simulations. In contrast to the earlier work \cite{Adlarson:2014ysb}, an extension to the GEANT3 software package \cite{geant3} provided by the authors of \cite{Bilger:2002aw} was used to simulate nuclear breakup of ${}^3\text{He}$ nuclei in the scintillator material. Additionally, the possibility that the primary proton-deuteron interaction does not occur with the pellet target but with evaporated target gas was accounted for. First, simulations of the signal reaction $pd\rightarrow {}^3\text{He}\,\eta$ were performed with $\cos\vartheta_{\eta}^*$ equally distributed over all values from $-1$ to $+1$. From this set of simulations, the product of acceptance times reconstruction efficiency was calculated as the ratio of the number of events reconstructed in a bin of $\cos\vartheta_{\eta}^*$ divided by the number of events that were generated in that bin. However, only if the detector resolution were perfect, would this ratio directly correspond to the sought-after product of acceptance and reconstruction efficiency. Otherwise, the finite detector resolution, in combination with angular distributions that exhibit a strong angular dependence, causes a bin migration effect in the opposite direction to the gradient of the angular distribution. In addition, the nuclear breakup introduces a tendency to reconstruct the ${}^3\text{He}$ nuclei at slightly smaller kinetic energies. To account for these effects, the acceptance correction is done in an iterative manner. For this, the angular distributions observed in data after correcting for the acceptance derived from the MC sample equally distributed in $\cos\vartheta_{\eta}^*$ are fitted by a third order polynomial
\begin{equation} f(\cos\vartheta_{\eta}^*) = N_0 \cdot \left(1+\alpha \cos\vartheta_{\eta}^* + \beta \cos^2\vartheta_{\eta}^* + \gamma \cos^3\vartheta_{\eta}^* \right) ~~ . \label{eq:pol}\end{equation}
These polynomials are subsequently used to generate a new set of MC simulations with which the product of acceptance and reconstruction efficiency can again be determined. This procedure is repeated until convergence of all angular distributions is reached. As an example, the angular distribution, along with the product of acceptance and reconstruction efficiency of the sum of the three measurements at $p_p = 1.70\; \text{GeV}/c$, is displayed in Fig \ref{fig:angdist-1700}.

\section{Normalization}

For the measurement presented here, normalization consists of two steps. The luminosity of the sum of the three measurements at $p_p = 1.70\; \text{GeV}/c$ ($Q_\eta=61.7\;\text{MeV}$) is determined by comparison of the integral over the fit to the ${}^3\text{He}\,\eta$ angular distribution displayed in Fig. \ref{fig:angdist-1700} and the total cross section value of $\sigma=(388.1\pm 7.2_{\text{stat.}})\;\text{nb}$ (with an additional $15\%$ normalization uncertainty), as measured by the ANKE collaboration at $Q_\eta=60\;\text{MeV}$ \cite{Rausmann:2009dn}.
\begin{figure}[!ht]
	\centering
	\begin{minipage}{0.46\textwidth}
		\includegraphics[width=1.0\textwidth]{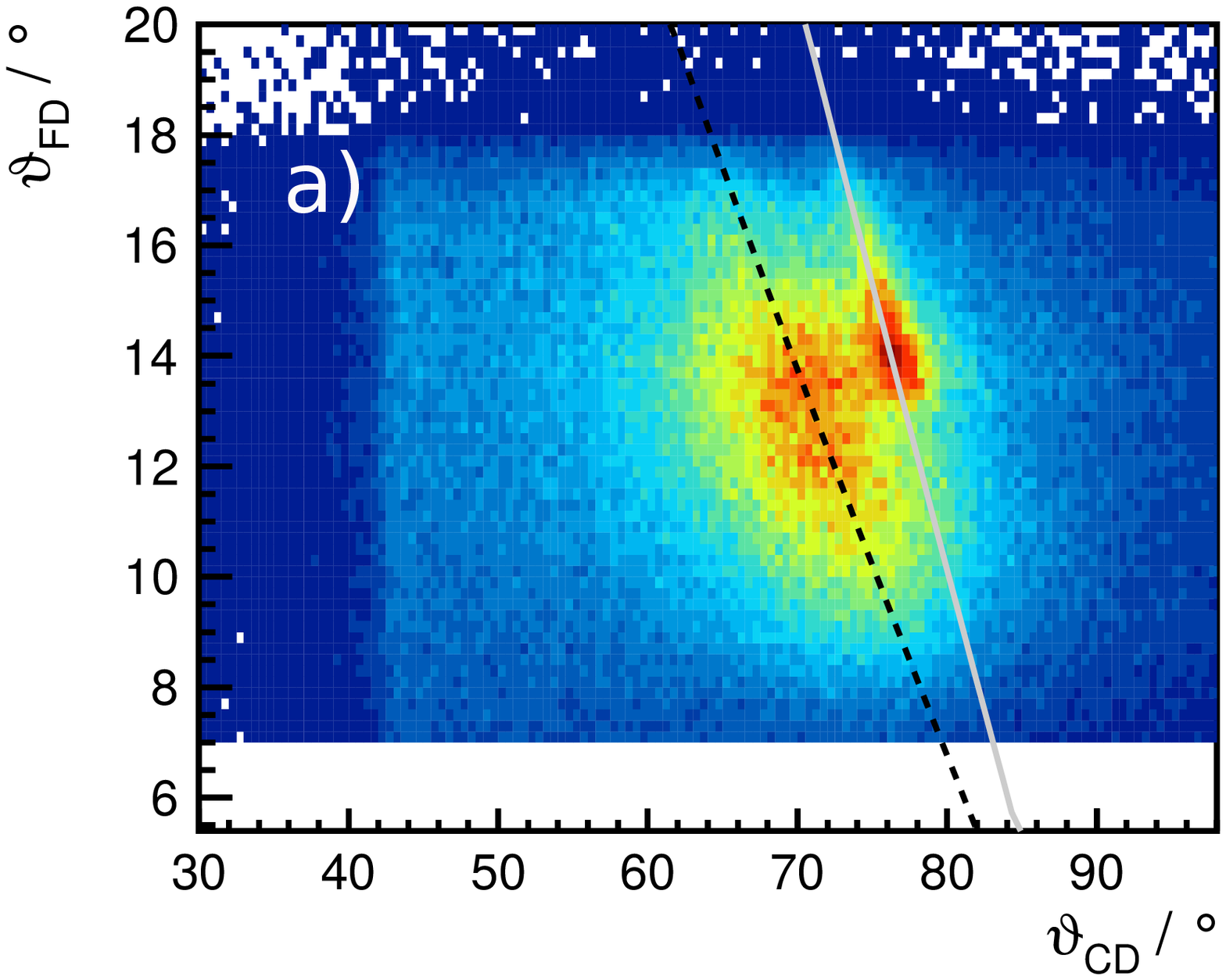}
	\end{minipage}
	\\
	\begin{minipage}{0.46\textwidth}
		\includegraphics[width=1.0\textwidth]{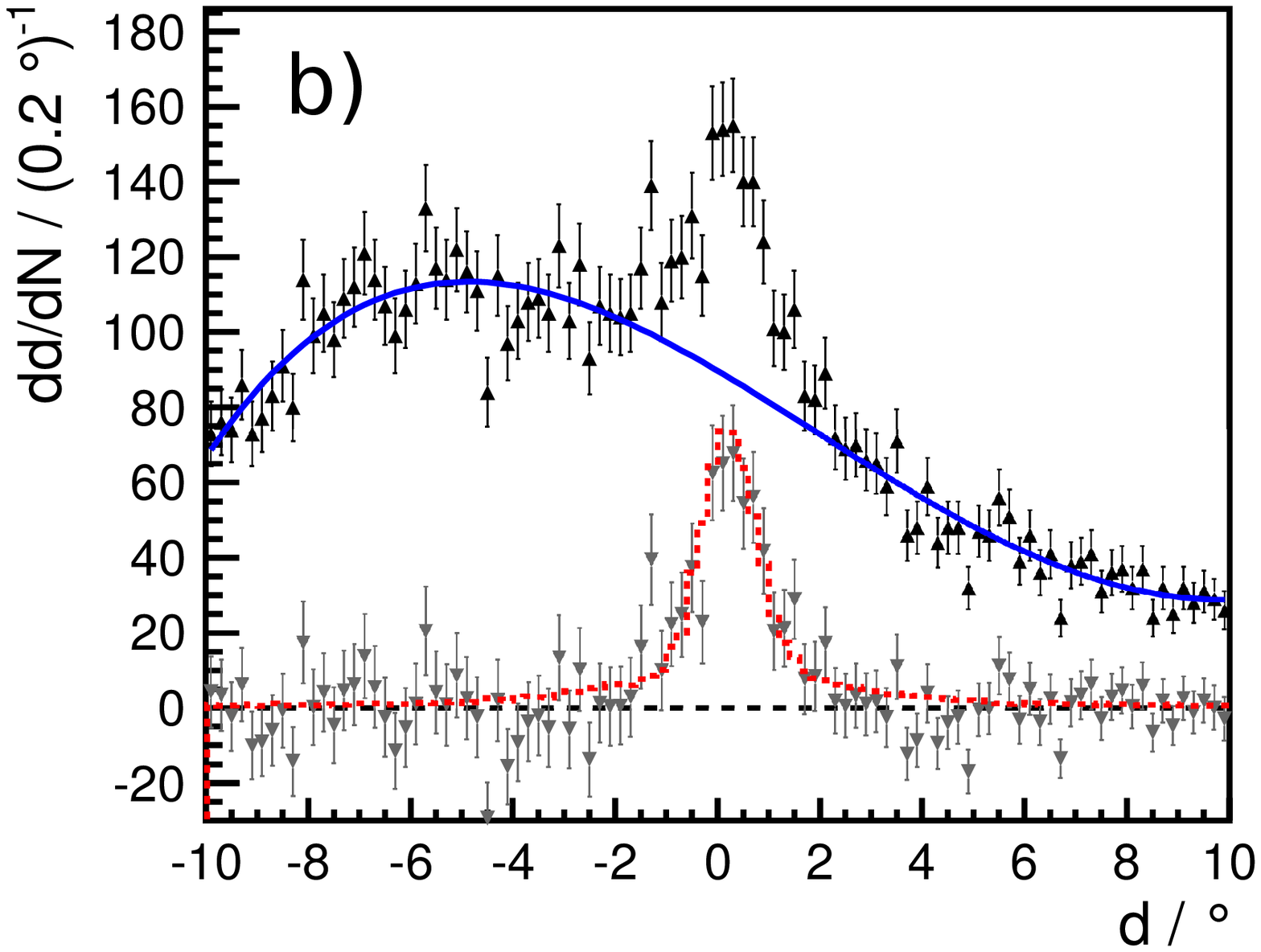}
	\end{minipage}
	\begin{minipage}{0.46\textwidth}
		\includegraphics[width=1.0\textwidth]{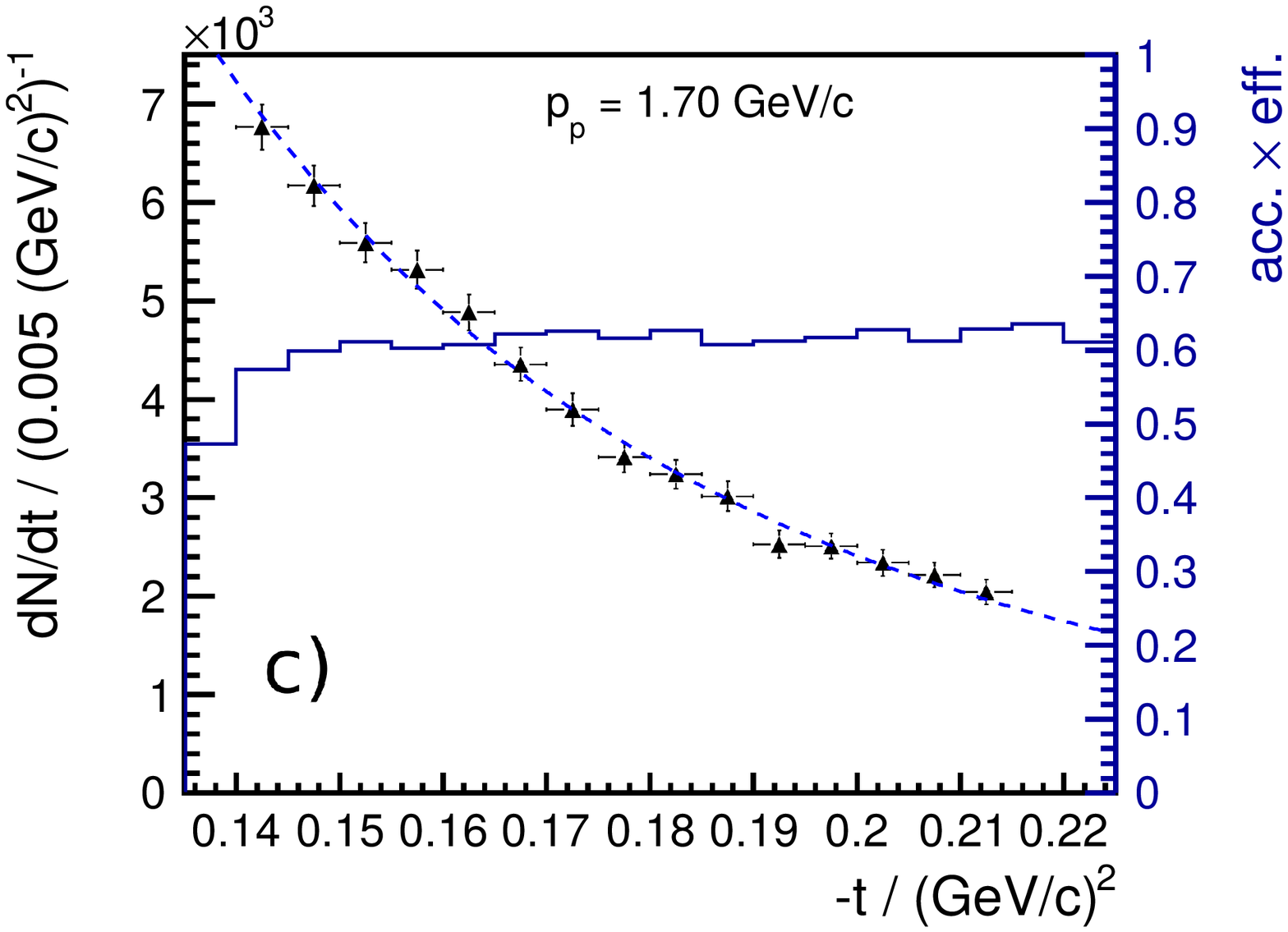}
	\end{minipage}
	\caption{{\bfseries a)} Pairs of polar angles of coincidently measured charged particles in the forward and central detectors, compared to the kinematical expectations for quasi-elastic $pd\rightarrow pp n_\text{spec}$ scattering (black dotted line) and $pd\rightarrow pd$ elastic scattering (gray line). {\bfseries b)} Projection onto the minimum distance $d$ of a given pair of polar angles to the kinematic relation for $pd$ elastic scattering, fitted by a fourth order polynomial (blue line). {\bfseries c)} Distribution of $pd$ elastic scattering events as a function of the momentum transfer $t$, fitted by a scaled fit to the literature data. The blue histogram represents the product of acceptance and reconstruction efficiency.}
	\label{fig:elas1}	
\end{figure}
Then, the measurements at the 14 remaining beam momenta are normalized relative to the luminosity derived for $p_p = 1.70\; \text{GeV}/c$. Whereas this relative normalization was performed using the single pion production $pd\rightarrow {}^3\text{He}\,\pi^0$ in \cite{Adlarson:2014ysb}, a different ansatz was used in this work. Here, the proton-deuteron elastic scattering is used for two reasons. On the one hand, data from \cite{Dalkhazhav:1969cma, Winkelmann:1980ca, Irom:1984wr, Velichko:1988ed, Guelmez:1991he} suggest that, within the experimental uncertainties, the $pd$ elastic differential cross section $d\sigma/dt$ does not vary with $p_p$, thus allowing one to cancel the literature cross section and its uncertainty in a relative normalization. On the other hand, as one of the objectives of this new measurement is to examine the cross section variation observed in \cite{Adlarson:2014ysb}, it is desirable to use an independent normalization method.\\
The elastic $pd$ scattering can be identified by demanding coincident charged particles in the forward and central detector. As the forward-going protons are minimum ionizing, the measurement of their energy deposit does not aid in determining their kinetic energy. Instead, the analysis is performed using only the measured scattering angles of the charged particles in the forward and central detector. First, a cut on the coplanarity of the two tracks is set at $120^\circ<|\varphi_{\text{FD}}-\varphi_{\text{CD}}|<240^\circ$. Afterwards, the polar angles of the two particles are compared. In the case of a two-particle final state, the polar angles of both particles are directly related and can each be expressed as a function of the other angle. In Fig. \ref{fig:elas1}a, this relation is displayed for data in comparison to the kinematical expectation. Under the assumption of an event corresponding to proton-deuteron elastic scattering, the momentum transfer $t$ is calculated as a function of the polar angle of the forward-going proton. In addition, the minimum distance $d$ to the kinematic expectation for $pd$ elastic scattering is calculated for each pair of measured polar angles $\vartheta_{\text{FD}}$ and $\vartheta_{\text{CD}}$. As can be seen from Fig. \ref{fig:elas1}b, the distance $d$ exhibits a narrow peak close to $d=0$ for momentum transfers in the region $0.140\;(\text{GeV}/c)^2\leq -t \leq 0.215\;(\text{GeV}/c)^2$ on top of a strong background contribution caused by quasi-elastic proton-proton scattering $pd\rightarrow pp n_\text{spec}$. For 15 bins in the momentum transfer range indicated above, the distribution of the distance $d$ is fitted by a fourth order polynomial for the background, again excluding the signal region. In this way, the acceptance-corrected event yield for proton-deuteron elastic scattering as a function of $-t$ is determined for each beam momentum (see Fig. \ref{fig:elas1}c). A fit to the combined database from \cite{Dalkhazhav:1969cma, Winkelmann:1980ca, Irom:1984wr, Velichko:1988ed, Guelmez:1991he}, given by $f(-t) = \exp(12.45 - 27.24\;(\text{GeV}/c)^{-2}\cdot |t| + 26.31\;(\text{GeV}/c)^{-4}\cdot |t|^2)$ originally performed by the ANKE collaboration and already applied for the luminosity determination in \cite{Mielke:2014xbu}, is scaled to the observed distribution $dN/dt$ in a one-parameter fit. The data used for the luminosity determination in \cite{Mielke:2014xbu} is currently being prepared for publication \cite{CFMPVK-PrivCom}. Under the assumption, discussed later, that the cross section $d\sigma/dt$ is independent of the beam momentum, the relative luminosity of a measurement at $p_p^\prime$ compared to the sum of the three measurements at $p_p=1.70\;\text{GeV}/c$ is directly given by the ratio of the two scaling factors. 

\section{Results}

\begin{figure*}[!ht]
	\centering
	\includegraphics[width=0.8\textwidth]{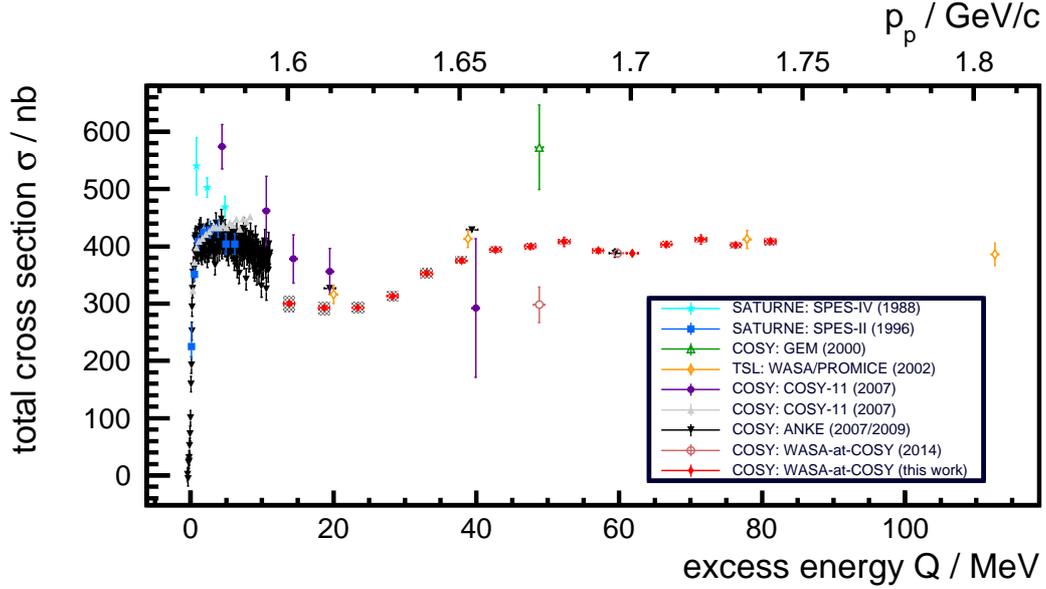}
	\caption{Total cross section of the reaction $pd\rightarrow {}^3\text{He}\,\eta$. Cyan stars are from \cite{Berger:1988ba}, blue boxes from \cite{Mayer:1995nu}, green open triangles from \cite{Betigeri:1999qa}, orange open diamonds from \cite{Bilger:2002aw}, purple filled circles from \cite{Smyrski:2007nu}, gray upward filled triangles from \cite{Adam:2007gz}, black downward filled triangles from \cite{Mersmann:2007gw, Rausmann:2009dn}, brown open circles from \cite{Adlarson:2014ysb} and red filled diamonds from the present work. Here, the error bars indicate the statistical point-to-point uncertainty, red boxes indicate the statistical chain-to-point uncertainty relative to the fixed cross section at $Q_\eta=61.7\;\text{MeV}$ and gray boxes indicate the systematic uncertainty. In addition, a normalization uncertainty of $16.3\%$ is not displayed here. Similarly, the normalization uncertainties of the literature data are not displayed.}
	\label{fig:totalcross}	
\end{figure*}

\begin{figure}[!ht]
	\centering
	\begin{minipage}{0.50\textwidth}
	\includegraphics[width=1.0\textwidth]{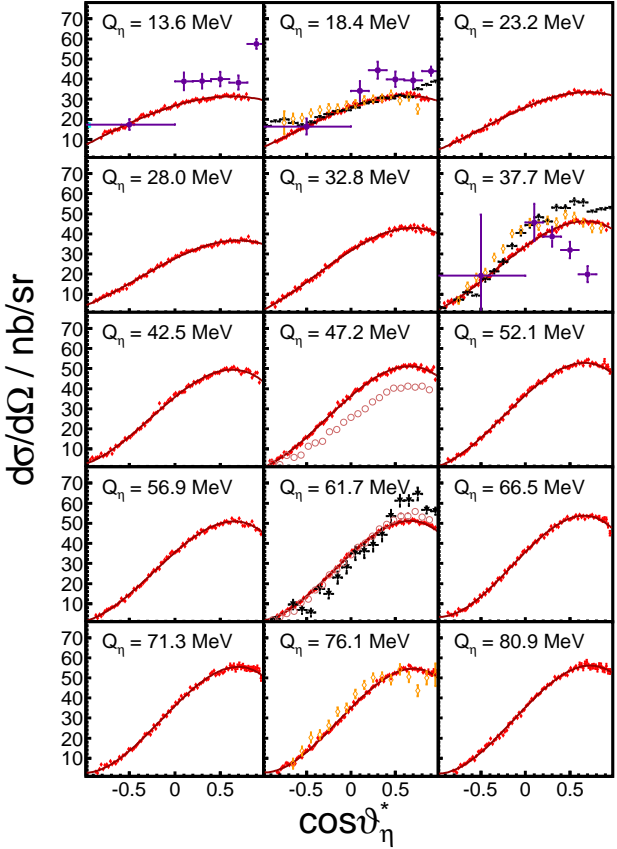}
	\caption{Differential cross sections of the reaction $pd\rightarrow {}^3\text{He}\,\eta$ at 15 excess energies between $Q_\eta=13.6\;\text{MeV}$ and $Q_\eta=80.9\;\text{MeV}$. The dark red line represents a fit of a third order polynomial as given in Eq.(\ref{eq:pol}). Data from the literature are shown for comparison, wherever possible. The color code is the same as in Fig. \ref{fig:totalcross}. The data from \cite{Betigeri:1999qa} were omitted due to the large uncertainties.}
	\label{fig:diffcross}	
	\end{minipage}
\end{figure}

\begin{figure}[!ht]
	\centering
	\begin{minipage}{0.46\textwidth}
	\includegraphics[width=1.0\textwidth]{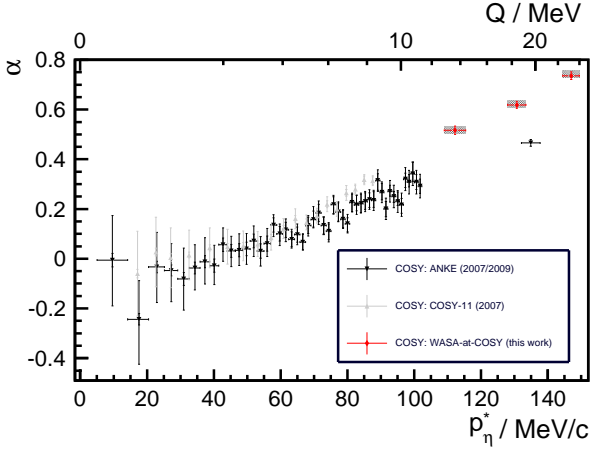}
	\caption{Asymmetry parameter $\alpha$ of the angular distributions of the reaction $pd\rightarrow {}^3\text{He}\,\eta$, comparing the presented data (red diamonds) to values extracted by the ANKE \cite{Mersmann:2007gw, Rausmann:2009dn} and COSY11 \cite{Smyrski:2007nu} experiments (black downward and gray upward triangles, respectively). Systematic uncertainties of the present work are shown as gray boxes. In the case of the data from \cite{Mersmann:2007gw, Rausmann:2009dn} and \cite{Smyrski:2007nu}, thick lines represent statistical uncertainties, thin lines systematic ones.}
	\label{fig:alpha}	
	\end{minipage}
\end{figure}

The resulting total cross sections at all 15 excess energies are given in table \ref{tab:totalcross} and displayed in Fig. \ref{fig:totalcross}, where they are compared to the database available in the literature. As the cross section at $Q_\eta=61.7\;\text{MeV}$ is fixed to the value of $\sigma=388.1\;\text{nb}$, the statistical uncertainty of the new measurement at that excess energy needs to be considered as a collective uncertainty of the whole chain of points relative to the fixed value. Additionally, asymmetric systematic uncertainties were found. Generally, using the supercycle mode of the accelerator, systematic effects due to changes to the experiment or environmental conditions can be ruled out. Also, by individual analysis and comparison of the three measurements at $p_p = 1.70\; \text{GeV}/c$, no systematic changes between the data-taking periods were found. Performing a relative normalization, systematic effects due to inefficiencies are largely canceled out. Two main sources of systematic uncertainty remain. The distribution and density of evaporated target gas in the scattering chamber is not known to high precision. As a shift of the vertex location along the beam axis leads to a loss of information for large polar angles $\vartheta$, variation of density and distribution in Monte Carlo simulations has implications on the geometrical acceptance that are larger for higher excess energies, when the maximum scattering angle of the ${}^3\text{He}$ nuclei is larger. In addition, while the assumption that the differential cross section of $pd$ elastic scattering $d\sigma/dt$ is constant as a function of the beam momentum is in accordance with the precision of the available data, calculations by \cite{Platonova:2016xjq, PlatPrivCom} have shown that the integral over the cross section $d\sigma/dt$ in the interval $0.140\;(\text{GeV}/c)^2\leq -t \leq 0.215\;(\text{GeV}/c)^2$ changes slightly but linearly with the beam momentum. Keeping the measured value at $p_p = 1.70\; \text{GeV}/c$ at a fixed position, the linear change in $d\sigma/dt$ introduces a change in normalization by roughly $4\%$ at $p_p = 1.60\; \text{GeV}/c$ and $-2\%$ at $p_p = 1.74\; \text{GeV}/c$. In addition, the overall normalization factor from the comparison of the $Q_\eta=61.7\;\text{MeV}$ data with the total cross section published in \cite{Rausmann:2009dn} comes with an uncertainty of $16.3\%$. Of this, $15\%$ is the normalization uncertainty of the literature cross section and an additional $6.3\%$ uncertainty was found when different subparts of the differential cross section were used for normalization instead of the total cross section. These $16.3\%$ are, however, irrelevant when the energy dependence of the total cross section is studied.
\begin{table}[!ht]
\centering
\caption{Total cross section of the reaction $pd\rightarrow {}^3\text{He}\,\eta$, including statistical point-to-point uncertainties $\Delta\sigma_{\text{stat}}^{P2P}$, the uncertainty of the whole dataset relative to the fixed point at $Q_\eta=61.7\;\text{MeV}$ $\Delta\sigma_{\text{stat}}^{C2P}$, and the systematic uncertainties $\Delta\sigma^\pm_{\text{sys}}$. The behaviour of the systematic uncertainty changes direction at $Q_\eta=61.7\;\text{MeV}$, as indicated by the sign. In addition, an overall normalization uncertainty of $16.3\%$ needs to be included.}
\label{tab:totalcross}
\begin{tabular}{cccccc}
$Q_\eta$ & $\sigma$ & $\Delta\sigma_{\text{stat}}^{P2P}$ & $\Delta\sigma_{\text{stat}}^{C2P}$ & $\Delta\sigma^-_{\text{sys}}$ & $\Delta\sigma^+_{\text{sys}}$ \\
in MeV & in nb & in nb & in nb & in nb & in nb\\ \hline
13.6(8) & 300.3 & 6.5 & 3.4 & -14.9 & 12.5 \\
18.4(8) & 292.2 & 5.8 & 3.3 & -11.8 & 11.0 \\
23.2(8) & 292.8 & 5.8 & 3.3 & -10.3 & 9.8 \\
28.0(8) & 312.9 & 6.0 & 3.5 & -8.1 & 9.3 \\
32.9(8) & 352.6 & 7.0 & 4.0 & -7.3 & 8.9 \\
37.7(8) & 374.7 & 7.3 & 4.2 & -4.3 & 8.0 \\
42.5(8) & 394.0 & 8.0 & 4.4 & -3.7 & 6.7 \\
47.3(8) & 399.8 & 7.6 & 4.5 & -2.8 & 5.1 \\
52.1(8) & 408.0 & 8.1 & 4.6 & -2.1 & 3.5 \\
56.9(8) & 392.7 & 7.2 & 4.4 & -0.1 & 1.7 \\
61.7(8) & 388.1 &  &  &  &   \\
66.5(8) & 403.3 & 7.8 & 4.5 & 2.6 & -1.8 \\
71.3(8) & 412.0 & 8.4 & 4.6 & 2.8 & -3.6 \\
76.1(8) & 402.5 & 7.7 & 4.5 & 3.3 & -5.4 \\
80.9(8) & 408.7 & 7.9 & 4.6 & 2.3 & -7.4  \\
\hline
\end{tabular} 
\end{table}
From Fig. \ref{fig:totalcross}, it is apparent that the sharp variation of the total cross section that was previously reported in \cite{Adlarson:2014ysb} is not confirmed by the present measurement. However, repeating the normalization procedure used in \cite{Adlarson:2014ysb}, it could be shown that the behaviour observed in \cite{Adlarson:2014ysb} is indeed reproduced. In \cite{mythesis}, it is shown in detail that the effect is caused by an incorrect assumption regarding the differential cross section of the single pion production.
In the excess energy interval $20\;\text{MeV} \lesssim Q_\eta \lesssim 60\;\text{MeV}$, the increase and subsequent leveling of the total cross section of the reaction $pd\rightarrow {}^3\text{He}\,\eta$ that was observed in \cite{Bilger:2002aw, Rausmann:2009dn} is also observed in the present work. It can, however, be studied in a lot more detail than was previously possible.\\
The differential cross sections derived in the present work are displayed in Fig. \ref{fig:diffcross}. Generally, the distributions at all energies exhibit the forward-peaking that was previously observed in other experiments. At all energies, the differential cross sections can be described by a third order polynomial, with no need for a quartic term. Due to the large amount of data gathered, the angular distributions as well as their energy dependence can be studied in unprecedented detail. The values of the fit parameters $N_0$, $\alpha$, $\beta$ and $\gamma$ of Eq.(\ref{eq:pol}) are given in tables \ref{tab:fitpar1} through \ref{tab:fitpar3} along with statistical and systematic uncertainties. Here, apart from the systematic uncertainty due to the aforementioned evaporated gas, an additional element arises from minor imprecisions in the determination of the polar angle. In relative normalization, this effect cancels and thus does not influence the determination of the total cross section. The asymmetry parameter $\alpha$ is of special importance, as it is regularly used to study an interference between $s$- and $p$-waves in the near-threshold data (see, e.g., \cite{Mersmann:2007gw, WILKIN200792}) in the search for indications of $\eta$-mesic states below threshold. In Fig. \ref{fig:alpha}, the three lowest energy values of the present work are compared to the values of the asymmetry parameter extracted in \cite{Mersmann:2007gw, Rausmann:2009dn} and \cite{Smyrski:2007nu}. Slightly better agreement with the higher values from \cite{Smyrski:2007nu} is found. The ANKE value at $Q=19.5\;\text{MeV}$ is in strong conflict to the findings reported here, however, as is already argued in \cite{Rausmann:2009dn}, the inclusion of this point into a combined fit with the data from \cite{Mersmann:2007gw} yields an unsatisfactory result.
\begin{table}[!ht]
	\centering
	\caption{Values of the fit parameter $N_0$ of the function given Eq.(\ref{eq:pol}) at all 15 excess energies.}
	\label{tab:fitpar1}
	\begin{tabular}{ccccc}
$Q_\eta$ & $N_0$ & $\Delta N_{0,\text{stat}}$ & $\Delta N^-_{0,\text{sys}}$ & $\Delta N^+_{0,\text{sys}}$ \\
in MeV & in nb/sr & in nb/sr & in nb/sr & in nb/sr\\ \hline
13.6(8) & 26.81 & 0.46 & 0.84 & 0.20 \\
18.4(8) & 26.22 & 0.40 & 0.54 & 0.15\\
23.2(8) & 25.96 & 0.40 & 0.30 & 0.15\\
28.0(8) & 27.72 & 0.44 & 0.17 & 0.10\\
32.9(8) & 31.68 & 0.58 & 0.17 & 0.17\\
37.7(8) & 33.78 & 0.64 & 0.21 & 0.51\\
42.5(8) & 35.77 & 0.74 & 0.14 & 0.62\\
47.3(8) & 36.29 & 0.71 & 0.14 & 0.72\\
52.1(8) & 36.72 & 0.77 & 0.14 & 0.67\\
56.9(8) & 35.49 & 0.67 & 0.14 & 0.84\\
61.7(8) & 34.71 & 0.63 & 0.12 & 0.75\\
66.5(8) & 35.68 & 0.72 & 0.13 & 0.96\\
71.3(8) & 36.02 & 0.78 & 0.12 & 0.94\\
76.1(8) & 35.03 & 0.70 & 0.13 & 0.97\\
80.9(8) & 35.29 & 0.72 & 0.18 & 0.83\\ \hline
	\end{tabular} 
\end{table}	
\begin{table}[!ht]
	\centering
	\caption{Values of the fit parameter $\alpha$ of the function given Eq.(\ref{eq:pol}) at all 15 excess energies.}
	\label{tab:fitpar2}
	\begin{tabular}{ccccc}
$Q_\eta$ & $\alpha$ & $\Delta \alpha_{\text{stat}}$ & $\Delta\alpha^-_{\text{sys}}$ & $\Delta\alpha^+_{\text{sys}}$ \\
in MeV &  &  &  & \\ \hline
13.6(8) & 0.517 & 0.017 & 0.012 & 0.015 \\
18.4(8) & 0.619 & 0.014 & 0.009 & 0.018 \\
23.2(8) & 0.736 & 0.015 & 0.009 & 0.022 \\
28.0(8) & 0.804 & 0.014 & 0.011 & 0.023 \\
32.9(8) & 0.894 & 0.014 & 0.008 & 0.026 \\
37.7(8) & 0.948 & 0.013 & 0.010 & 0.023 \\
42.5(8) & 1.025 & 0.014 & 0.008 & 0.022 \\
47.3(8) & 1.054 & 0.013 & 0.008 & 0.026 \\
52.1(8) & 1.101 & 0.013 & 0.009 & 0.027 \\
56.9(8) & 1.118 & 0.013 & 0.007 & 0.022 \\
61.7(8) & 1.183 & 0.008 & 0.009 & 0.023 \\
66.5(8) & 1.253 & 0.014 & 0.009 & 0.022 \\
71.3(8) & 1.257 & 0.014 & 0.008 & 0.017 \\
76.1(8) & 1.285 & 0.014 & 0.008 & 0.020 \\
80.9(8) & 1.306 & 0.015 & 0.008 & 0.017 \\ \hline
	\end{tabular} 
\end{table}	
\begin{table}[!ht]
	\centering
	\caption{Values of the fit parameters $\beta$ and $\gamma$ of the function given Eq.(\ref{eq:pol}) at all 15 excess energies. Systematic uncertainties are omitted here and can be found in \cite{mythesis}.}
	\label{tab:fitpar3}
	\begin{tabular}{ccccc}
$Q_\eta$ & $\beta$ & $\Delta \beta_{\text{stat}}$ & $\gamma$ & $\Delta \gamma_{\text{stat}}$ \\
in MeV &  &  &  &   \\ \hline
13.6(8)	&	-0.326	&	0.016	&	-0.098	&	0.041	\\
18.4(8)	&	-0.339	&	0.012	&	-0.180	&	0.028	\\
23.2(8)	&	-0.307	&	0.012	&	-0.213	&	0.030	\\
28.0(8)	&	-0.305	&	0.012	&	-0.255	&	0.028	\\
32.9(8)	&	-0.343	&	0.011	&	-0.296	&	0.027	\\
37.7(8)	&	-0.352	&	0.011	&	-0.356	&	0.026	\\
42.5(8)	&	-0.371	&	0.011	&	-0.463	&	0.026	\\
47.3(8)	&	-0.370	&	0.010	&	-0.438	&	0.024	\\
52.1(8)	&	-0.347	&	0.011	&	-0.480	&	0.025	\\
56.9(8)	&	-0.358	&	0.010	&	-0.511	&	0.024	\\
61.7(8)	&	-0.331	&	0.010	&	-0.560	&	0.016	\\
66.5(8)	&	-0.302	&	0.011	&	-0.652	&	0.026	\\
71.3(8)	&	-0.269	&	0.012	&	-0.599	&	0.027	\\
76.1(8)	&	-0.257	&	0.012	&	-0.624	&	0.029	\\
80.9(8)	&	-0.235	&	0.013	&	-0.605	&	0.031	\\ \hline
	\end{tabular} 
\end{table}	
\section{Summary}

In the course of this work, total and differential cross sections of the $\eta$ meson production in proton-deuteron fusion were extracted. The differential distributions exhibit the same forward-peaking behaviour as previously observed away from the reaction threshold. Due to the amount and quality of the data, it is possible for the first time to study changes in the shape of the angular distributions with rising excess energy in a large interval between $13.6\;\text{MeV}$ and $80.9\;\text{MeV}$. In this way, the contributions of higher partial waves might be studied which will greatly aid in the investigation of the production process that remains largely unknown.\\
A previously reported sharp variation of the total cross section around $Q_\eta\approx 50\;\text{MeV}$ is not confirmed. However, the fluctuating structure of the production cross section between $Q_\eta\approx 10\;\text{MeV}$ and $Q_\eta\approx 60\;\text{MeV}$ that had already been observed by both the WASA/PROMICE and ANKE experiments \cite{Bilger:2002aw, Rausmann:2009dn}, albeit in much less detail, is nicely reproduced. Close to the production threshold, effects of a strong final state interaction are thought to be a dominating contribution to the total cross section. The observed structure reported here might indicate the energy region in which the final state interaction loses its importance. With none of the available theoretical models being able to reproduce the forward-peaking in the angular distributions as well as the observed total cross section, further theoretical effort is clearly needed in order to fully understand the production of $\eta$ mesons off ${}^3\text{He}$ nuclei. 

\section*{Acknowledgements}

The present work received funding from the European Union Seventh Framework Programme (FP7/2007-2013) under grant agreement number 283286. We gratefully acknowledge the support given by the Forschungszentrum J\"ulich FFE Funding Program of the J\"ulich Center for Hadron Physics, by the Polish National Science Centre through the grant No. 2016/23/B/ST2/00784, and by the DFG through the Research Training Group GRK2149. We thank the COSY crew for their work and the excellent conditions during the beam time and Dr. M. N. Platonova and Dr. V. I. Kukulin for their valuable contributions regarding the proton-deuteron elastic scattering. 


\bibliography{elsarticle-he3eta.bib}

\end{document}